\documentclass[aps,pra,twocolumn]{revtex4}
\usepackage[a4paper,top=1.8cm,bottom=1.8cm,left=1.8cm,right=1.8cm]{geometry}
\usepackage{graphicx}
\usepackage{amsmath,amssymb}
\usepackage{dcolumn}
\usepackage{epsfig}
\newcommand{\ket}[1]{\ensuremath{\left|#1\right>}}
\newcommand{\bra}[1]{\ensuremath{\left<#1\right|}}
\newcommand{\mf}[1]{\boldsymbol{#1}}
\newcommand{\mc}[1]{\ensuremath{\mathcal{#1}}}

\def\email{\footnote{benhurcris@yahoo.com}}
\def\correspondingauthor{\footnote{Corresponding author: rarun@cutn.ac.in}}
\begin{document}
\title{Squeezing in resonance fluorescence via vacuum induced coherences}
\author{H. B. Crispin\email{}}
\author{R. Arun\correspondingauthor{}}
\affiliation{Department of Physics, School of Basic $\&$ Applied Sciences, Central University of
	Tamilnadu, Thiruvarur 610005, Tamilnadu, India.}
\date{\today}
\begin{abstract}
The squeezing spectrum of the fluorescence field emitted from a four-level atom in $J=1/2$ to $J=1/2$ configuration
driven by two coherent fields is studied. We find that the squeezing properties of the fluorescence radiation are
significantly influenced by the presence of vacuum-induced coherence in the atomic system. It is shown that such
coherence induces spectral squeezing in phase quadratures of the fluorescence light for both weak and strong driving
fields. The dependence of the squeezing spectrum on the relative phase of the driving fields is also investigated.
Effects such as enhancement or suppression of the squeezing peaks are shown in the spectrum as the relative
phase is varied. An analytical explanation of the numerical findings is presented using dressed-states of the
atom-field system.
\end{abstract}
\maketitle
\newpage
\section{\label{sec:intro}INTRODUCTION}
\vspace{-1.5em}
Resonance fluorescence has played a significant role in understanding the essentials of the interaction between light and matter
\cite{moll, kim, mand, sqslush}. Many interesting phenomena demonstrating the quantum features of light, such as photon antibunching
\cite{kim}, sub-Poissonian photon statistics \cite{mand},  and squeezing \cite{sqslush} have been experimentally observed.
Squeezed light, a non-classical state of radiation, has quantum fluctuations in one of its quadrature components reduced below its
shot-noise limit. Squeezed states of radiation have been thoroughly investigated over the last few decades due to its theoretical
and practical significance \cite{walls, yuen}. Among the many works in the literature related to squeezing, the phenomena of
squeezing in resonance fluorescence was explored by many authors
\cite{walls2, zoller, ficek, jakob, zhou, vog, dal, fic, fic2, gao, gao1}. These works considered either the total variance of the
fluorescence field in a selected phase quadrature or the spectral components of phase quadratures, i.e., squeezing spectrum, to study
squeezing in atomic fluorescence \cite{walls2, zoller}. Walls and Zoller were the first to show theoretically that a driven two-level
atom exhibits total variance squeezing in fluorescence radiation \cite{walls2}. They also found that for weak excitation the
out-of-phase quadrature noise spectrum shows squeezing at the laser frequency \cite{zoller}. Ficek and Swain reported large
fluorescence-squeezing in a coherently driven two-level system coupled to squeezed vacuum \cite{ficek}. Theoretical studies also
demonstrated squeezing characteristics dependent on nonlinear two-photon emission processes \cite{jakob} and tunable two-mode
squeezing in the fluorescence of two-level systems \cite{zhou}. Studies extended to three-level atoms showed that atomic coherences,
decay rates of the atomic transitions and Rabi frequencies play an important role in modifying the squeezing spectra
\cite{vog, dal, fic, fic2, gao}. Further, Gao \textit{et al.} predicted that ultranarrow squeezing peaks may appear in the
squeezing spectrum of a coherently driven $V$-type atom \cite{gao1}.

Another important phenomenon that has been discussed extensively in the context of resonance fluorescence is the effect
of vacuum-induced coherence. It is well understood that even in the absence of any external driving field, there can be coherence
between near-degenerate atomic states decaying via common vacuum modes \cite{inter1,inter2,inter3,ficek2,inter4,anton1}.
This type of coherence induced by the vacuum field is known as vacuum-induced coherence (VIC). The VIC leads to many remarkable
effects in the fluorescence \cite{inter1,inter2} and absorption \cite{inter3} properties of atomic media. An excellent review of
the study of VIC effects in multilevel atoms is given by Ficek and Swain \cite{ficek2}. The role of VIC has also been
investigated in the phase-dependent squeezing spectra of driven atoms \cite{inter4,anton1}. It was shown that the VIC gives rise
to a strong enhancement or broadening of squeezing in the spectrum over a wide range of parameters \cite{inter4}. Gonzalo
{\it et al.} \cite{anton1} have shown that the VIC in driven $\Lambda$-type systems may induce spectral squeezing in phase quadratures
of the fluorescence in contrast to the usual situation where VIC is not included.

In all these publications, it is assumed that the dipole moments of the allowed atomic-transitions are non-orthogonal for the
VIC to exist in decay processes. This condition is not favorable to the experimental realization in real atomic systems. Many
alternative schemes were suggested to circumvent this problem \cite{bypass}. In an interesting paper, Kiffner \textit{et al.}
proposed a novel scheme to observe VIC in the atomic fluorescence \cite{kif1}. They considered a four-level atom with
$J=1/2$ to $J=1/2$ transition which is driven by a linearly polarized light \cite{kif1, kif2}. Since the dipole moments of
the $\pi$ transitions in this atomic model are antiparallel (non-orthogonal), this system is a good candidate to probe
for VIC effects. This has motivated further studies such as squeezing spectrum \cite{tan} and collective resonance fluorescence
\cite{collect} in this system. In these studies \cite{kif1,kif2,tan,collect}, the fluorescence properties are investigated
when the $\pi$ transitions in the atom are driven by a linearly polarized light. Recently, we have shown that the effect of VIC
in the fluorescence spectrum becomes stronger when an additional $\sigma^{-}$-polarized light drives the atom with
$J=1/2$ to $J=1/2$ transition \cite{heb}. It has been shown that the incoherent spectrum of fluorescence emitted on the
$\sigma$ transitions is dependent on the relative phase of the applied fields. Thus, it would be interesting to see how
VIC affects the squeezing spectra and to study the phase control of squeezing in this system. Therefore, in this paper,
considering the same arrangement as in \cite{heb}, we study the influence of VIC in the squeezing spectrum and examine how
the relative phase of the driving fields alters the squeezing properties of the fluorescence radiation. The atomic model
considered in this study can be realized experimentally using ${}^{198}\text{Hg}^+$ ions \cite{eich} in an optical trap.
The investigation of squeezing spectrum in this model system is thus a realistic approach to probe for VIC effects in
fluorescence which have remained elusive in the much studied $V$-type \cite{inter4} and $\Lambda$-type \cite{anton1}
three-level atoms considered in previous works.  We show many interesting features such as splitting of squeezing peaks,
ultranarrow squeezing peaks, strong reduction in quantum fluctuations for both weak and strong-driving fields due to VIC,
and phase-controlled enhancement and cancellation of spectral squeezing.

Our paper is organized as follows. In section \ref{sec:atom}, we discuss the atomic model and present the basic density matrix equations
describing the interaction of the atom with the driving fields. In section \ref{sec:sqspec}, we derive the formulas for the squeezing spectra
of the fluorescence emitted on the $\pi$ and $\sigma$ transitions. The numerical results and their dressed-state interpretation are given in
section \ref{sec:numres}. Finally, section \ref{sec:concl} presents a summary of the main results.\vspace{-1.5em}
\begin{figure}[t]
\begin{center}
\includegraphics[width=8cm,height=5cm]{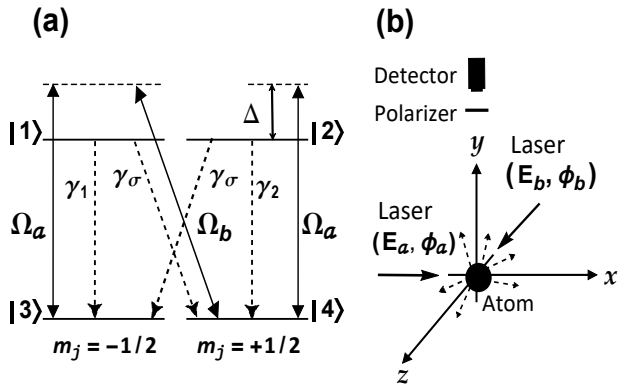}
\caption{(a) The level scheme of a driven four-level atom with $J=1/2$ to $J=1/2$ transitions. A linearly polarized
field drives the transitions $\ket{1}\leftrightarrow\ket{3}$ and $\ket{2}\leftrightarrow\ket{4}$ while a
$\sigma^{-}$-polarized field couples to the transition $\ket{1}\leftrightarrow\ket{4}$ in the atom. (b) The
arrangement for atom-laser interaction and the detection of the fluorescence light.}
\end{center}
\vspace{-1.8em}
\end{figure}
\section{\label{sec:atom}THE HAMILTONIAN AND DENSITY MATRIX EQUATIONS}
\vspace{-1em}
The system of interest consists of a four-level atom with $J=1/2$ to $J=1/2$ transition as shown in figure 1(a). This level scheme
has a doubly degenerate excited and ground atomic states with energy separation $\hbar \omega_{o}$. Spontaneous emission causes an atom in the excited
levels ($\ket{1}$ and $\ket{2}$) to decay to both ground levels ($\ket{3}$ and $\ket{4}$). The transitions $\ket{1}\leftrightarrow\ket{3}$ and
$\ket{2}\leftrightarrow\ket{4}$ in the atom have antiparallel (real) dipole moments and will be referred to as the $\pi$ transitions. The cross
transitions $\ket{1}\leftrightarrow\ket{4}$ and $\ket{2}\leftrightarrow\ket{3}$ in the atom have orthogonal (complex) dipole moments and are
designated as $\sigma$ transitions. The dipole moments for these allowed transitions can be obtained from the matrix elements of the dipole
moment operator $\mf{\hat{d}}$. By use of the Wigner-Eckart theorem \cite{sak}, they are calculated as
\begin{eqnarray}
\mf{d}_{1}&=&\bra{1}\mf{\hat{d}}\ket{3}=\frac{-1}{\sqrt{3}}\,\mc{D}\,\mf{e}_{z},~~ \mf{d}_{2}=\bra{2}\mf{\hat{d}}\ket{4}=-\mf{d}_{1},\nonumber\\
\mf{d}_{3}&=&\bra{2}\mf{\hat{d}}\ket{3}=\sqrt{\frac{2}{3}}\,\mc{D}\,\mf{\epsilon^{(-)}},~\mf{d}_{4} = \bra{1}\mf{\hat{d}}\ket{4}=\mf{d}_{3}^{*},
\label{matel}
\end{eqnarray}
where $\mf{\epsilon^{(-)}}=(\mf{e}_{x}-i\mf{e}_{y})/\sqrt{2}$ is the circular polarization vector and $\mc{D}$ represents the reduced dipole
matrix element of the operator $\mf{\hat{d}}$.

We assume that the atomic system is interacting with two coherent fields propagating in perpendicular directions. The applied fields have equal
frequencies and drive the atom in a setup shown in figure 1(b). A linearly polarized light (amplitude $E_a$, phase $\phi_a$, polarization
$\mf{e}_{z}$) is assumed to propagate along the x-direction and drives the $\pi$ transitions ($\ket{1}\leftrightarrow\ket{3}$,
$\ket{2}\leftrightarrow\ket{4}$) in the atom. In addition, a circularly polarized light (amplitude $E_b$, phase $\phi_b$, polarization
$\mf{\epsilon^{(-)}}$) traveling along the z-direction couples the $\sigma$ transition $\ket{1}\leftrightarrow\ket{4}$ in the atom.
Using the rotating-wave and electric-dipole approximations, the atom-field Hamiltonian of the system can be written as
\begin{flalign}
H=&~\hbar\omega_{o}(A_{11}+A_{22})+\hbar[\Omega_{a}(A_{13}-A_{24})e^{-i(\omega_{l}t+\phi_{a})}&\nonumber\\
&-\Omega_{b}A_{14}e^{-i(\omega_{l}t+\phi_{b})}+h.c.],&
\end{flalign}
where $\omega_{l}$ is the frequency of the applied fields, $\omega_{o} =\omega_{13}=\omega_{24}$ is the resonance frequency of the
atomic transitions, $\Omega_{a}=\mc{D}\,E_{a}/(\sqrt{3}\hbar)$ is the Rabi frequency of the linearly polarized light, and
$\Omega_{b}=\sqrt{2}\mc{D}\,E_{b}/(\sqrt{3}\hbar)$ is the Rabi frequency of the circularly polarized light driving the atom. The
operators $A_{lk}=\ket{l}\bra{k}$ represent the atomic transitions for $l\neq k$ and the populations for $l=k$.

The dynamics of the atom-field interaction and the spontaneous emission processes can be described using the master equation for the
density operator. For convenience, we use an interaction picture by making the unitary transformation
\begin{equation}
\text{U}= \exp\{i[\omega_{l}t+\phi_{a}](A_{11}+A_{22})+i[\phi_{a}-\phi_{b}](A_{22}+A_{44})\}.  \nonumber
\end{equation}
In this interaction picture, the Hamiltonian becomes independent of time and the phases $(\phi_a, \phi_b)$ of the
applied fields.  The interaction picture Hamiltonian is given by
\begin{flalign}
H_{I}=&-\hbar\Delta(A_{11}+A_{22})&\nonumber\\&+\hbar(\Omega_{a}(A_{13}-A_{24})-\Omega_{b}A_{14}+h.c.),\label{hamil} &
\end{flalign}
where $\Delta=\omega_{l}-\omega_{o}$ corresponds to the detuning of the driving fields from resonance with the atomic transitions.

With the inclusion of spontaneous decay terms, the time evolution of the density matrix elements can be obtained using equation (\ref{hamil})
to be \cite{heb}
\begin{flalign}
\dot{\rho}_{11}=&-(\gamma_{1}+\gamma_{\sigma})\rho_{11}+i\Omega_{a}(\rho_{13}-\rho_{31})-i\Omega_{b}(\rho_{14}-\rho_{41}),&
\label{rho1}
\end{flalign}
\begin{flalign}
\dot{\rho}_{33}=&~\gamma_{1}\rho_{11}+\gamma_{\sigma}\rho_{22}-i\Omega_{a}(\rho_{13}-\rho_{31}),&
\label{rho2}
\end{flalign}
\begin{flalign}
\dot{\rho}_{44}=&~\gamma_{\sigma}\rho_{11}+\gamma_{2}\rho_{22}+i\Omega_{a}(\rho_{24}-\rho_{42})+i\Omega_{b}(\rho_{14}-\rho_{41}),&
\label{rho3}
\end{flalign}
\begin{flalign}
\dot{\rho}_{12}=&~\left(\!-\frac{(\gamma_{1}+\gamma_{2})}{2}-\gamma_{\sigma}\!\right)\rho_{12}-i\Omega_{a}(\rho_{32}+\rho_{14})&\nonumber\\
&+i\Omega_{b}\rho_{42},&
\label{rho4}
\end{flalign}
\begin{flalign}
\dot{\rho}_{13}=&~\left(\!-\frac{(\gamma_{1}+\gamma_{\sigma})}{2}+i\Delta\!\right)\rho_{13}+i\Omega_{a}(\rho_{11}-\rho_{33})&\nonumber\\
&+i\Omega_{b}\rho_{43},&
\label{rho5}
\end{flalign}
\begin{flalign}
\dot{\rho}_{23}=&~\left(\!-\frac{(\gamma_{2}+\gamma_{\sigma})}{2}+i\Delta \!\right)\rho_{23}+i\Omega_{a}(\rho_{21}+\rho_{43}),&
\label{rho6}
\end{flalign}
\begin{flalign}
\dot{\rho}_{14}=&~\left(\!-\frac{(\gamma_{1}+\gamma_{\sigma})}{2}+i\Delta \!\right)\rho_{14}-i\Omega_{a}(\rho_{12}+\rho_{34})&\nonumber\\
&-i\Omega_{b}(\rho_{11}-\rho_{44}),&
\label{rho7}
\end{flalign}
\begin{flalign}
\dot{\rho}_{24}=&~\left(\!-\frac{(\gamma_{2}+\gamma_{\sigma})}{2}+i\Delta \!\right)\rho_{24}-i\Omega_{a}(\rho_{22}-\rho_{44})&\nonumber\\
&-i\Omega_{b}\rho_{21},&
\label{rho8}
\end{flalign}
\begin{flalign}
\dot{\rho}_{34}=&~\gamma_{12}\rho_{12}-i\Omega_{a}(\rho_{32}+\rho_{14})-i\Omega_{b}\rho_{31}.&
\label{rho9}
\end{flalign}
In equations (\ref{rho1})-(\ref{rho9}), $\gamma_{1}=\gamma_{2}=\gamma/3$ denotes the decay rates of the $\pi$ transitions and
$\gamma_{\sigma}=2\gamma/3$ is the decay rate of the $\sigma$ transitions (see figure 1(a)). The total decay rate of each of the excited
atomic states is given by $\gamma=\gamma_{1}+\gamma_{\sigma}=\gamma_{2}+\gamma_{\sigma}$. The $\gamma_{12}$-term is responsible for
VIC effects in the atom. It can be written as $\gamma_{12} = (\mf{d_{1}}\cdot\mf{d_{2}^*}/|\mf{d_{1}}||\mf{d_{2}}|)
\sqrt{\gamma_{1}\gamma_{2}}= -\sqrt{\gamma_{1}\gamma_{2}}$, which arises because the transitions $\ket{1}\leftrightarrow\ket{3}$ and
$\ket{2}\leftrightarrow\ket{4}$ undergo spontaneous emission via common vacuum modes \cite{kif1,kif2}. Note that $\gamma_{12}$ is negative
since the dipole moments $\mf{d_{1}}$ and $\mf{d_{2}}$ are antiparallel. If $\gamma_{12}$ is taken to be zero $(\gamma_{12}=0)$, then
VIC effect will be absent in spontaneous emission.

To solve for the steady-state dynamics of the driven atom, we eliminate $\rho_{22}$ using the trace condition $(\hbox{Tr}\rho=1)$ and
rewrite equations (\ref{rho1})-(\ref{rho9}) in a simplified form
\begin{equation}
\frac{d}{dt}\mf{\hat{\psi}(t)}=\hat{M}\,\mf{\hat{\psi}(t)}+\hat{C},\label{stdyeq}
\end{equation}
where  $\hat{M}$ is a $15\times15$ matrix whose elements are the coefficients in equations (\ref{rho1})-(\ref{rho9}) with $\mf{\hat{\psi}}$
being a $15\times1$ column vector of density matrix elements
\begin{align}
\mf{\hat{\psi}} = &\left(\langle A_{11}\rangle,\langle A_{33}\rangle,\langle A_{44}\rangle,\langle A_{12}\rangle,\langle A_{21}\rangle,
\langle A_{13}\rangle, \langle A_{31}\rangle,\right.&\nonumber\\&\left.\langle A_{23}\rangle,\langle A_{32}\rangle,\langle A_{14}\rangle,
\langle A_{41}\rangle,\langle A_{24}\rangle,\langle A_{42}\rangle,\langle A_{34}\rangle,\langle A_{43}\rangle\right)^{T}\!\!\!.\label{psicol}
\end{align}
Here $\langle A_{ij}\rangle=\rho_{ji}$ and the inhomogeneous term $\hat{C}$ in equation (\ref{stdyeq}) is also a $15\times1$ column vector
with non-zero components $\hat{C}_{2}=\gamma_{\sigma},\hat{C}_{3}=\gamma_{2},\hat{C}_{12}=i\Omega_{a},\hat{C}_{13}=-i\Omega_{a}$. In steady state, the
behavior of the system can be described by the solution of equation (\ref{stdyeq}) in the long-time limit $\mf{\hat{\psi}}(\!\infty\!)=
-\hat{M}^{-1}\,\hat{C}$, which is obtained by setting $d\hat{\psi}/dt=0$. The stead-state values of the density matrix elements are found to be
\begin{flalign}
\rho_{11} = \rho_{22}
&~=\frac{4\Omega_{a}^{4}}{2\Omega_{a}^{2}(\gamma^{2}+4\Delta^{2}+8\Omega_{a}^{2})+\Omega_{b}^{2}(\gamma^{2}+4\Delta^{2})},&\nonumber
\end{flalign}
\begin{equation}
\rho_{33}=\frac{4\Omega_{a}^{4}+(\Omega_{a}^{2}+\Omega_{b}^{2})(\gamma^{2}+4\Delta^{2})}{2\Omega_{a}^{2}(\gamma^{2}+4\Delta^{2}+8\Omega_{a}^{2})
+\Omega_{b}^{2}(\gamma^{2}+4\Delta^{2})},\nonumber
\end{equation}
\begin{equation}
\rho_{44}=\frac{\Omega_{a}^{2}(\gamma^{2}+4\Delta^{2}+4\Omega_{a}^{2})}{2\Omega_{a}^{2}(\gamma^{2}+4\Delta^{2}+8\Omega_{a}^{2})+\Omega_{b}^{2}
(\gamma^{2}+4\Delta^{2})},
\nonumber
\end{equation}
\begin{flalign}
\rho_{13}= - \rho_{24} &~=\frac{4\Omega_{a}^{3}(\Delta-i\gamma/2)}{2\Omega_{a}^{2}(\gamma^{2}+4\Delta^{2}+8\Omega_{a}^{2})+\Omega_{b}^{2}
(\gamma^{2}+4\Delta^{2})},&\nonumber
\end{flalign}
\begin{equation}
\rho_{23}=\frac{-4\Omega_{a}^{2}\Omega_{b}(\Delta-i\gamma/2)}{2\Omega_{a}^{2}(\gamma^{2}+4\Delta^{2}+8\Omega_{a}^{2})+\Omega_{b}^{2}(
\gamma^{2}+4\Delta^{2})}, \nonumber  \label{sted}
\end{equation}
\begin{equation}
\rho_{34}=\frac{\Omega_{a}\Omega_{b}(\gamma^{2}+4\Delta^{2})}{2\Omega_{a}^{2}(\gamma^{2}+4\Delta^{2}+8\Omega_{a}^{2})+\Omega_{b}^{2}
(\gamma^{2}+4\Delta^{2})},\nonumber
\end{equation}
\begin{equation}
\rho_{12} = \rho_{14} = 0. \label{sted}
\end{equation}
It is clear from equations (\ref{sted}) that steady-state populations and coherences  have no dependence on the VIC parameter ($\gamma_{12}$)
and the phases ($\phi_{a}$, $\phi_{b}$) of the driving fields. However, as was already shown earlier \cite{kif1,kif2,heb}, the fluorescence
properties of the atomic system depend strongly on $\gamma_{12}$ and the relative phase $(\phi_a-\phi_b)$ of the driving fields.
\vspace{-1em}
\section{\label{sec:sqspec}CALCULATION OF THE SQUEEZING SPECTRUM}
\vspace{-1em}
We now turn to derive analytic expressions useful for the calculation of squeezing spectra of the fluorescence fields. In our system,
the fluorescence field is composed of light emanating from both the $\pi$ and $\sigma$ transitions in the atom. The electric field
operator of the source field can be decomposed into a sum of positive and negative frequency parts as $\vec{E}(\vec{r},t)
= \vec{E}^{(+)}(\vec{r},t) + \vec{E}^{(-)}(\vec{r},t)$. In the far-field zone, they are given by \cite{scull}
\begin{eqnarray}
\vec{E}^{(+)}_{\pi}(t)&=&-\frac{\omega_{o}^2}{c^2r}[\{\mf{\hat{r}}\times(\mf{\hat{r}}\times\mf{\vec{d}}_{31})\}A_{31}(t')\nonumber\\
&&+\{\mf{\hat{r}}\times (\mf{\hat{r}}\times\mf{\vec{d}}_{42})\}A_{42}(t')]e^{-i(\omega_{l}t'+\phi_{a})},\nonumber\\
\vec{E}^{(+)}_{\sigma}(t)&=&-\frac{\omega_{o}^2}{c^2r}[\{\mf{\hat{r}}\times(\mf{\hat{r}}\times\mf{\vec{d}}_{41})\}A_{41}(t')\nonumber\\
&&+\{\mf{\hat{r}}\times (\mf{\hat{r}}\times\mf{\vec{d}}_{32})\}\nonumber \\
&&\times A_{32}(t')e^{-2i(\phi_{a}-\phi_{b})}] e^{-i(\omega_{l}t'+\phi_{b})},\nonumber\\
\vec{E}^{(-)}_{\pi}(t)&=&~\left[\vec{E}^{(+)}_{\pi}(t)\right]^{\dagger},\nonumber\\
\vec{E}^{(-)}_{\sigma}(t)&=&~\left[\vec{E}^{(+)}_{\sigma}(t)\right]^{\dagger}.\label{+elec}
\end{eqnarray}
Here $t'=t-r/c$ is the retarded time and $\mf{\hat{r}}=\vec{\mf{r}}/r$ is the unit vector along the observation point
$\vec{\mf{r}}$. The subindex $\pi$ ($\sigma$) in equations (\ref{+elec}) refers to the fluorescence field of the $\pi$ $(\sigma)$
transitions. We choose the direction of observation ($\mf{\hat{r}}$) of the fluorescence to be along the y-direction,
as shown in figure 1(b). It is seen from equations (\ref{+elec}) that the light coming from the $\pi$ transitions will be linearly
polarized along $\mf{e}_{z}$ and the light emitted from the $\sigma$ transitions will have polarization along $\mf{e}_{x}$. Thus,
using a polarization filter to differentiate the fluorescence fields, the squeezing spectra of the fluorescence from $\pi$ and
$\sigma$ transitions can be studied separately.

In squeezing measurements done via homodyne detection schemes, the fluorescence field (signal field) is superimposed with a
local oscillator field (a reference field having a controllable phase and same frequency as the laser frequency $\omega_l$)
and the intensity correlation of the superposed fields is measured \cite{ou}. In this setup, the two-time correlation of a
field-quadrature is the quantity of interest. The slowly varying quadrature components of the electric field operator with
phase $\theta$ are defined by
\begin{align}
\vec{E}_{\pi}(\theta,t)=\vec{E}^{(+)}_{\pi}(t)e^{i(\omega_{l}t+\theta)}+\vec{E}^{(-)}_{\pi}(t)
e^{-i(\omega_{l}t+\theta)},\nonumber\\
\vec{E}_{\sigma}(\theta,t)=\vec{E}^{(+)}_{\sigma}(t)e^{i(\omega_{l}t+\theta)}+\vec{E}^{(-)}_{\sigma}(t)
e^{-i(\omega_{l}t+\theta)}.\label{elec2}
\end{align}
The spectrum of squeezing is defined in terms of normal-ordered correlation of the quadrature components as \cite{ou,knoll}
\begin{flalign}
S^{\pi}(\omega,\theta)=&\frac{1}{2 \pi}\int_{-\infty}^{\infty}\!\!d\tau e^{i\omega\tau}\hat{T}\langle:\!
\vec{E}_{\pi}(\theta,t),\vec{E}_{\pi}(\theta,t+\tau)\!:\rangle,& \nonumber \\
S^{\sigma}(\omega,\theta)=&\frac{1}{2 \pi}\int_{-\infty}^{\infty}\!\!d\tau e^{i\omega\tau}\hat{T}\langle:\!
\vec{E}_{\sigma}(\theta,t),\vec{E}_{\sigma}(\theta,t+\tau)\!:\rangle,& \label{spec}
\end{flalign}
where $\langle \vec{A},\vec{B}\rangle=\langle\vec{A}.\vec{B}\rangle-\langle \vec{A}\rangle.\langle \vec{B}\rangle$ and
$\hat{T}$ denotes the time ordering operator.

In the steady-state limit ($t\rightarrow\infty$), the two-time averages appearing in equations (\ref{spec}) can be easily
calculated using the quantum regression theorem \cite{lax} and the density matrix elements (\ref{sted}) . For this purpose,
we introduce column vectors of correlation functions
\vspace{-1.2em}
\begin{align}
\hat{U}^{mn}(t,\tau)&=& \nonumber \\
&\left[\langle \delta A_{11}(t+\tau)\delta A_{mn}(t)\rangle,\langle \delta A_{33}(t+\tau)\delta A_{mn}(t)\rangle,\right.&\nonumber
\\&\left.\langle \delta A_{44}(t+\tau)\delta A_{mn}(t)\rangle,\langle \delta A_{12}(t+\tau)\delta A_{mn}(t)\rangle,\right.&\nonumber\\
&\left.\langle \delta A_{21}(t+\tau)\delta A_{mn}(t)\rangle,\langle \delta A_{13}(t+\tau)\delta A_{mn}(t)\rangle,\right.&\nonumber\\
&\left.\langle \delta A_{31}(t+\tau)\delta A_{mn}(t)\rangle,\langle \delta A_{23}(t+\tau)\delta A_{mn}(t)\rangle,\right.&\nonumber\\
&\left.\langle \delta A_{32}(t+\tau)\delta A_{mn}(t)\rangle,\langle \delta A_{14}(t+\tau)\delta A_{mn}(t)\rangle,\right.&\nonumber\\
&\left.\langle \delta A_{41}(t+\tau)\delta A_{mn}(t)\rangle,\langle \delta A_{24}(t+\tau)\delta A_{mn}(t)\rangle,\right.&\nonumber\\
&\left.\langle \delta A_{42}(t+\tau)\delta A_{mn}(t)\rangle,\langle \delta A_{34}(t+\tau)\delta A_{mn}(t)\rangle,\right.&\nonumber\\
&\left.\langle \delta A_{43}(t+\tau)\delta A_{mn}(t)\rangle\right]^{T}\!\!\!,&\nonumber\\
&\left.m,n=1,2,3,4.\right. \label{2tim}&
\end{align}
\begin{figure}[t]
	\begin{center}
		\includegraphics[width=8cm,height=5.5cm]{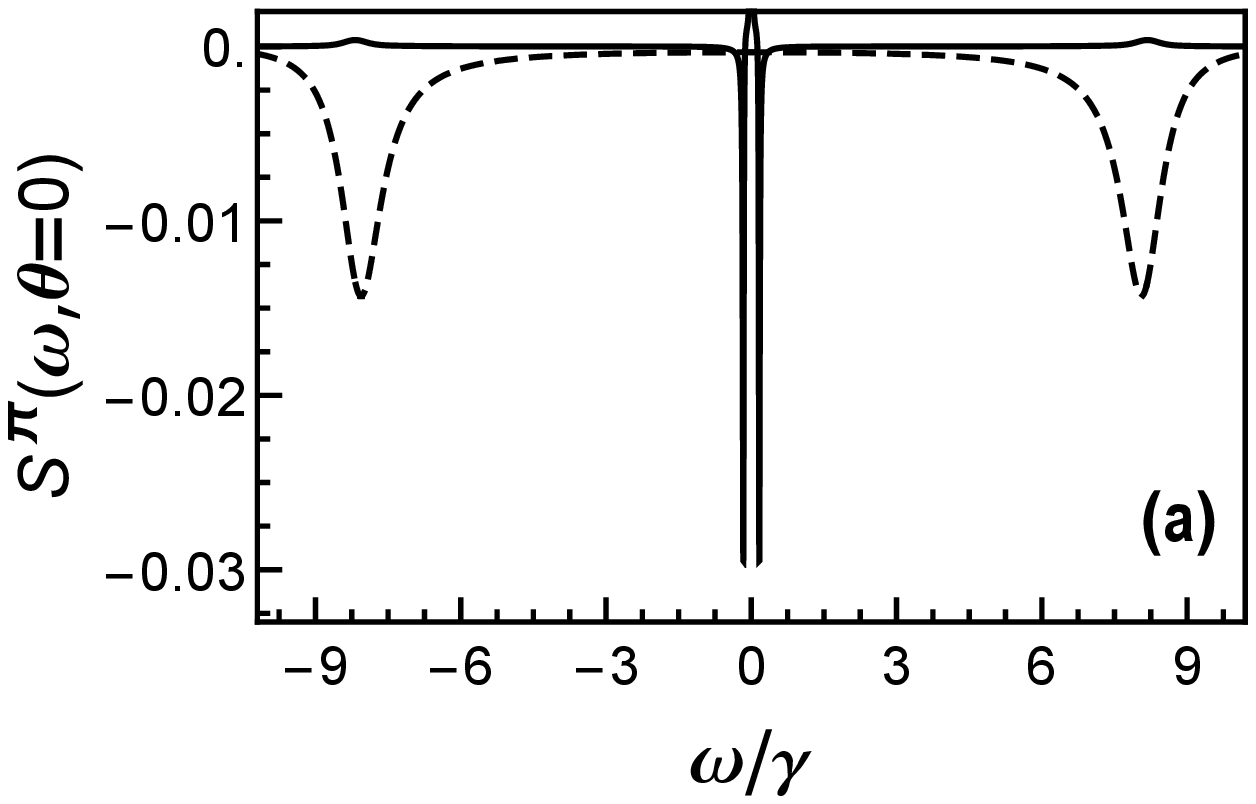}
	\end{center}
	\begin{center}
		\includegraphics[width=8cm,height=5.5cm]{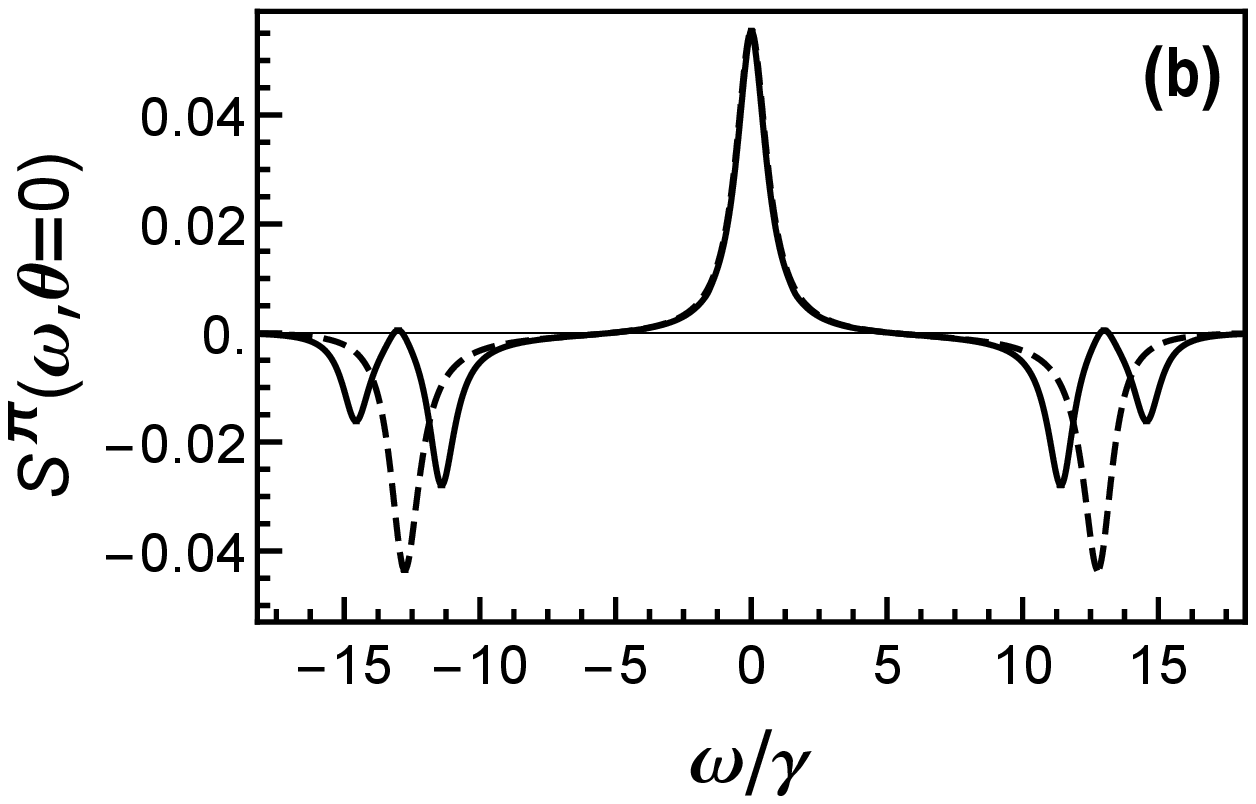}
	\end{center}
	\vspace{-2em}
\caption{Influence of the additional field on the squeezing spectra $S^{\pi}(\omega,\theta)$ as a function of $\omega$ for $\theta=0$, $\gamma=1$,
$\Delta=8$ and (a) $\Omega_{a}=0.6$, $\Omega_{b}=0.9$ and (b) $\Omega_{a}=5$, $\Omega_{b}=2$. The dashed curves are the spectra in the absence of the additional field ($\Omega_{b}=0$) with other parameters remaining same. Actual values of the dashed curves in (a) are 0.25 times that shown. Note that $\gamma_{12}=-\sqrt{\gamma_{1}\gamma_{2}}=-\gamma/3$ throughout the paper unless stated otherwise.}
\end{figure}
In this equation, $\delta A_{ij}(t)=A_{ij}(t)-\langle A_{ij} \rangle_{st}$ are the fluctuations of the atomic operators from their
steady-state mean values. According to the quantum regression theorem \cite{lax}, the two-time column vector $\hat{U}^{mn}(t,\tau)$
satisfies
\begin{equation}
\frac{d}{d\tau}\hat{U}^{mn}(t,\tau)=\hat{M}\hat{U}^{mn}(t,\tau),\label{2timdiff}
\end{equation}
where $\hat{M}$ is the $15\times15$ matrix defined in equation (\ref{stdyeq}). Now, following the method depicted in \cite{anton1}
for the time ordering of the operators in equations (\ref{spec})and applying the quantum regression theorem, the
squeezing spectra for the fluorescence fields of the $\pi$ and $\sigma$ transitions can be expressed as
\begin{flalign}
S^{\pi}(\omega,\theta)=&~f_{\pi}\textit{Re}\bigg\{\sum_{j=1}^{15}\displaystyle{\lim_{t \to \infty}}\left[\big\{\gamma_{1}\,
\hat{N}_{7,j}\,\hat{U}_{j}^{31}(t,0)\right.&\nonumber\\
&~+\gamma_{2}\,\hat{N}_{13,j}\,\hat{U}_{j}^{42}(t,0)+\gamma_{12}\,\hat{N}_{7,j}\,\hat{U}_{j}^{42}(t,0)&\nonumber\\
&~+\gamma_{12}\,\hat{N}_{13,j}\,\hat{U}_{j}^{31}(t,0)\big\}e^{2i(\theta-\phi_{a}+\omega_{l}r/c)}&\nonumber\\
&~+\gamma_{1}\,\hat{N}_{6,j}\,\hat{U}_{j}^{31}(t,0)+\gamma_{2}\,\hat{N}_{12,j}\,\hat{U}_{j}^{42}(t,0)&\nonumber\\
&~\left.+\gamma_{12}\,\hat{N}_{6,j}\,\hat{U}_{j}^{42}(t,0)+\gamma_{12}\,\hat{N}_{12,j}\,\hat{U}_{j}^{31}(t,0)\right]
\bigg\},&\label{pisq}
\end{flalign}
\begin{flalign}
S^{\sigma}(\omega,\theta)=&~f_{\sigma}\gamma_{\sigma}\textit{Re}\bigg\{\sum_{j=1}^{15}\displaystyle{\lim_{t \to \infty}}
\left[\big\{\hat{N}_{11,j}\,\hat{U}_{j}^{41}(t,0)\right.&\nonumber\\
&~+e^{-2i\phi}\,\hat{N}_{11,j}\,\hat{U}_{j}^{32}(t,0)+e^{-2i\phi}\hat{N}_{9,j}\,\hat{U}_{j}^{41}(t,0)&\nonumber\\
&~+e^{-4i\phi}\,\hat{N}_{9,j}\,\hat{U}_{j}^{32}(t,0)\big\}e^{2i(\theta-\phi_{b}+\omega_{l}r/c)}&\nonumber\\
&~+\hat{N}_{10,j}\,\hat{U}_{j}^{41}(t,0)+e^{-2i\phi}\,\hat{N}_{10,j}\,\hat{U}_{j}^{32}(t,0)&\nonumber\\
&~\left.+e^{2i\phi}\,\hat{N}_{8,j}\,\hat{U}_{j}^{41}(t,0)+\hat{N}_{8,j}\,\hat{U}_{j}^{32}(t,0)\right]\bigg\},&\label{sigsq}
\end{flalign}
where $\hat{N}_{i,j}$ represents the $(i,j)$ element of the matrix $\hat{N}=[(i \omega-\hat{M})^{-1}+(-i \omega-\hat{M})^{-1}]$
and $\phi=\phi_{a}-\phi_{b}$ is the relative phase of the applied fields. In equations (\ref{pisq}) and (\ref{sigsq}), the terms
$f_{\pi}$ and $f_{\sigma}$ are common prefactors which will be set to unity in the following. Note that the spectra
$S^{\pi}(\omega,\theta)$ and $S^{\sigma}(\omega,\theta)$ have an explicit dependence on the VIC term ($\gamma_{12}$) and
the relative phase $(\phi)$, respectively.

\vspace{-0.5em}
\section{\label{sec:numres}NUMERICAL RESULTS and dressed-state explanation}
\vspace{-1em}
In this section we discuss the numerical results of the squeezing spectra and then provide an explanation using dressed-states
for interpretation of the results. According to the criterion for squeezing \cite{ou}, a fluorescence field is squeezed in the frequency components
if the squeezing spectrum $S(\omega,\theta)$ in a selected quadrature ($\theta$) becomes negative for some frequencies $(\omega)$. To demonstrate
squeezing in the spectra $S^{\pi}(\omega,\theta)$ and $S^{\sigma}(\omega,\theta)$, we analyze the numerical results obtained using equations
(\ref{pisq}) and (\ref{sigsq}). In the numerical calculations, we assume $e^{2i(\omega_{l}r/c-\phi_{a})}=e^{2i(\omega_{l}r/c-\phi_{b})}=1$ and
scale all the parameters such as decay rates, detuning, and Rabi frequencies in units of $\gamma$.
\vspace{-1.5em}
\begin{figure}[t]
	\begin{center}
		\includegraphics[width=8cm,height=5.5cm]{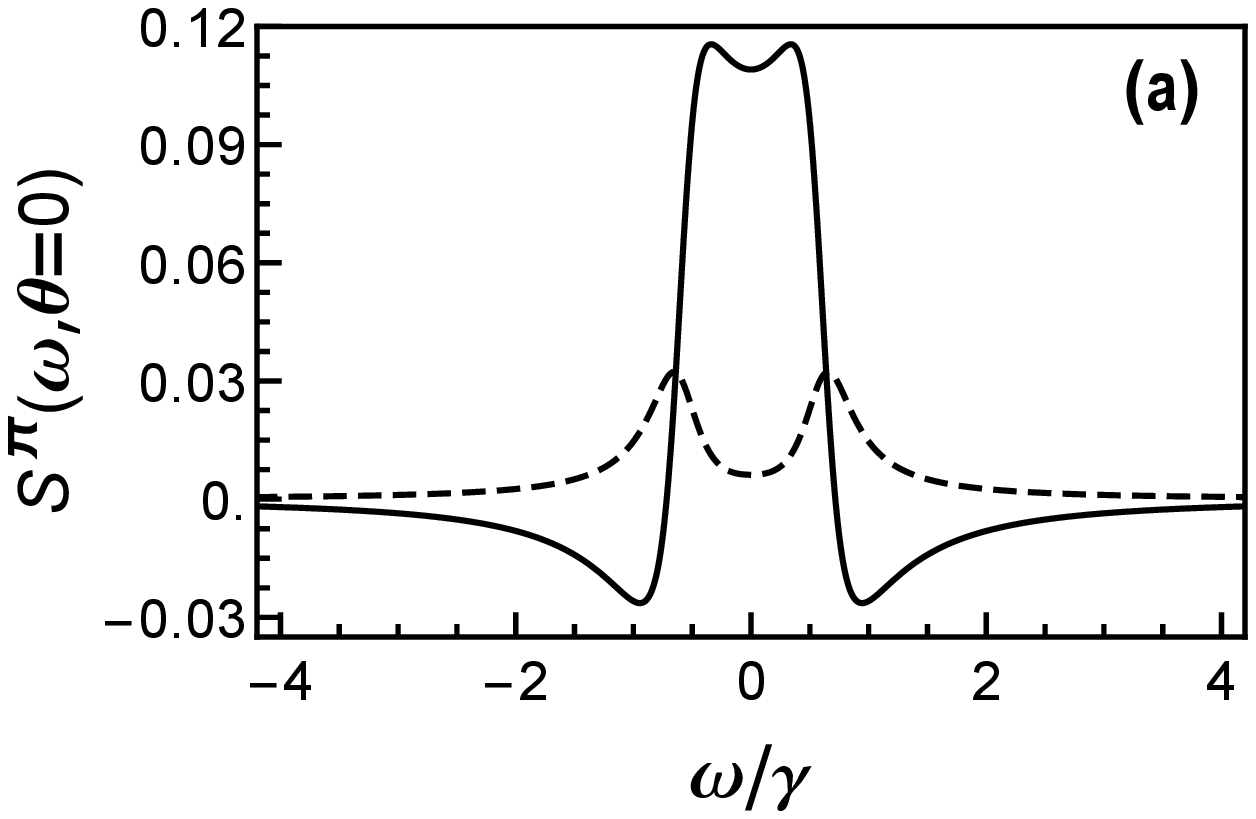}
	\end{center}
	\begin{center}
		\includegraphics[width=8cm,height=5.5cm]{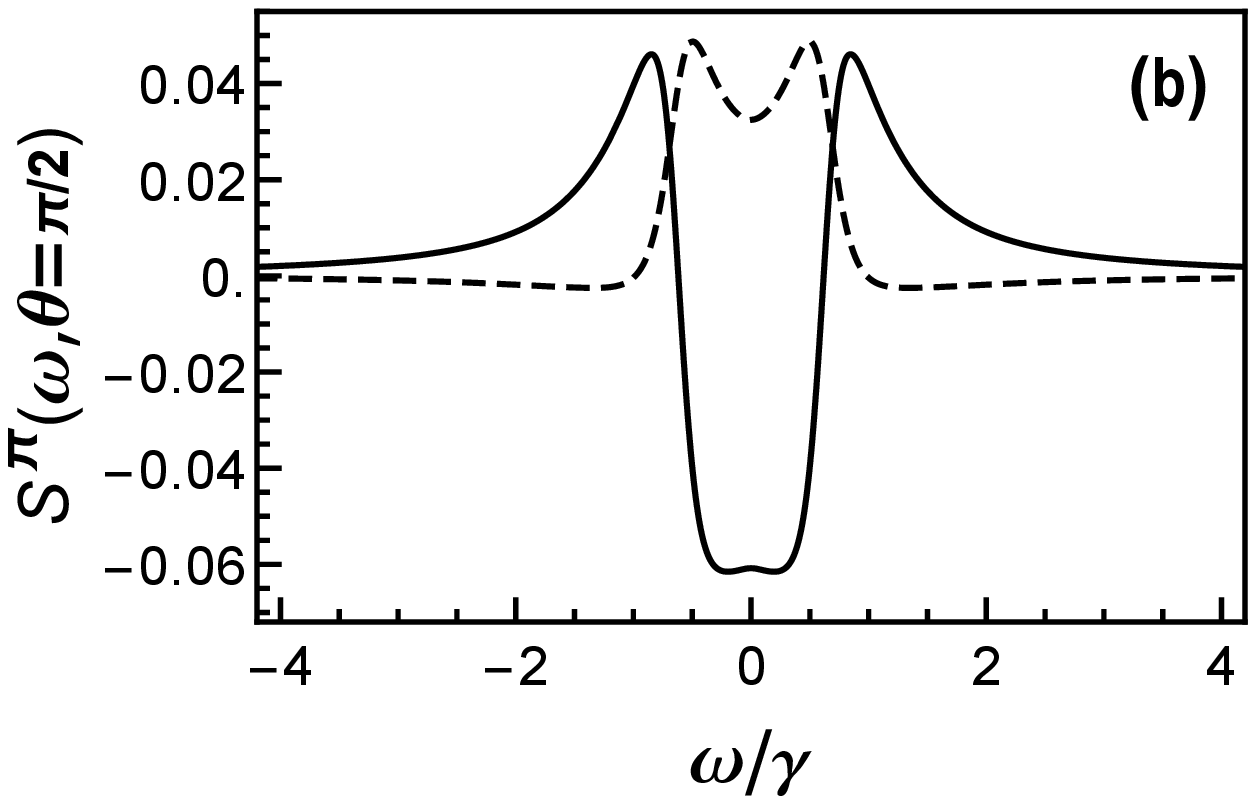}
	\end{center}
	\vspace{-1.5em}
	\caption{Squeezing spectra $S^{\pi}(\omega,\theta)$ as a function of $\omega$ for weak resonance excitation. The parameters are $\gamma=1$,
$\Delta=0$, $\Omega_{a}=0.2$,  $\Omega_{b}=0.6$, and (a) $\theta=0$ and (b) $\theta=\pi/2$. The solid (dashed) curves are in the presence (absence)
of VIC terms.}
\vspace{-1em}
\end{figure}
\subsection{Squeezing spectrum - $\pi$ transitions}
\vspace{-1em}
\subsubsection{Effect of the additional field}
\vspace{-1em}
We first consider the spectral squeezing in the fluorescence generated by the $\pi$ transitions in the atom. In figure 2, the numerical results are shown
for the in-phase quadrature spectra $[S^{\pi}(\omega,\theta=0)]$ when considering both weak and strong driving fields. The results are also compared with
the case when there is no additional $\sigma^{-}$-polarized light driving the atom. In the absence of the additional field $(\Omega_b=0)$, the squeezing
spectrum $S^{\pi}(\omega,\theta)$ is the same as that of a two-level atom as reported by Tan {\it et al.} \cite{tan}. As shown by the dashed
curves in figure 2, the spectrum $S^{\pi}(\omega,0)$ exhibits two-mode squeezing (for $\Omega_b=0$) at the Rabi sideband frequencies
$\Omega'=\pm\sqrt{4\Omega_{a}^2+\Delta^{2}}$, which is the well-known feature of a driven two-level atom for off-resonance excitations
\cite{zhou}. Interesting new features appear in spectral squeezing when the additional field $(\Omega_b\neq0)$ drives the atom. For weak driving
fields $(\Omega_a,\Omega_b<\gamma)$, ultranarrow squeezing peaks $(S^{\pi}(\omega,0)<0)$ are induced around the laser frequency $(\omega=0)$
as shown by the solid curve in figure 2(a). In the case of strong-field excitations $(\Omega_a,\Omega_b\gg\gamma)$, the additional field leads to
splitting of the squeezing peaks in the spectrum (compare solid and dashed curves in figure 2(b)).
\vspace{-1.5em}
\subsubsection{Influence of VIC}
\vspace{-1em}
We now consider the effect of VIC in the squeezing spectrum emitted from the $\pi$ transitions in the atom. In order to do that, we compare
the spectra with and without VIC terms. In the absence of VIC effects, the spectrum is obtained by substituting $\gamma_{12}=0$ in equations
(\ref{stdyeq}) and (\ref{pisq}). Note that all the terms having $\gamma_{12}$ in equation (\ref{pisq}) arise from the coupling
between the atomic transitions $\ket{3}\leftrightarrow\ket{1}$ and $\ket{4}\leftrightarrow\ket{2}$ and such coupling will be absent for
$\gamma_{12}=0$. Thus, $\pi$-fluorescence squeezing spectrum without considering VIC can be written as
\begin{flalign}
S^{\pi}(\omega,\theta)=&~\textit{Re}\bigg\{\sum_{j=1}^{15}\displaystyle{\lim_{t \to \infty}}\left[\big\{\gamma_1 \hat{N}_{7,j}\,
\hat{U}_{j}^{31}(t,0)\right.&\nonumber\\
&~\left.+\gamma_2 \hat{N}_{13,j}\,\hat{U}_{j}^{42}(t,0)\big\} e^{2i(\theta-\phi_{a}+\omega_{l}r/c)}\right.&\nonumber\\
&~\left.+\gamma_1 \hat{N}_{6,j}\,\hat{U}_{j}^{31}(t,0)+\gamma_2 \hat{N}_{12,j}\,\hat{U}_{j}^{42}(t,0)\right]\bigg\}.&\label{pinovic}
\end{flalign}
In figure 3 we show the squeezing spectra in two quadratures $(\theta=0,\pi/2)$ for the case of weak driving fields
($\Omega_{a},\Omega_{b}<\gamma$) on resonance. It is clear that in the absence of VIC $(\gamma_{12}=0)$ there is no squeezing
neither in the in-phase $(\theta=0$) nor in the out-of-phase $(\theta=\pi/2)$ quadrature spectrum
(see dashed curves in figure 3). The spectral features are greatly modified when considering the effect of VIC
$(\gamma_{12}=-\sqrt{\gamma_{1}\gamma_{2}})$ in the analysis. As shown by the solid curves in figure 3, the squeezing appears
in both quadratures as a consequence of VIC. This result is in contrast to that of
Tan \textit{et al.} \cite{tan} where VIC induces squeezing only in the out-of-phase quadrature for weak resonance excitations.

In previous studies concerning VIC effects, the fluorescence field was shown to exhibit spectral squeezing only for weak to moderately
intense driving fields \cite{inter4,anton1,tan}. However, we find that the VIC induces significant squeezing
in the present system even for strong driving fields ($\Omega_{a},\Omega_{b}>>\gamma$). This feature is illustrated in figure 4
where the quadrature spectra are given for strong (off-resonance) driving fields. It is evident that squeezing is obtained
because of VIC in both the in-phase and out-of-phase quadratures although at different frequencies (compare solid and dotted
curves in figure 4).

\begin{figure}[b]
	\begin{center}
		\includegraphics[width=8cm,height=5.5cm]{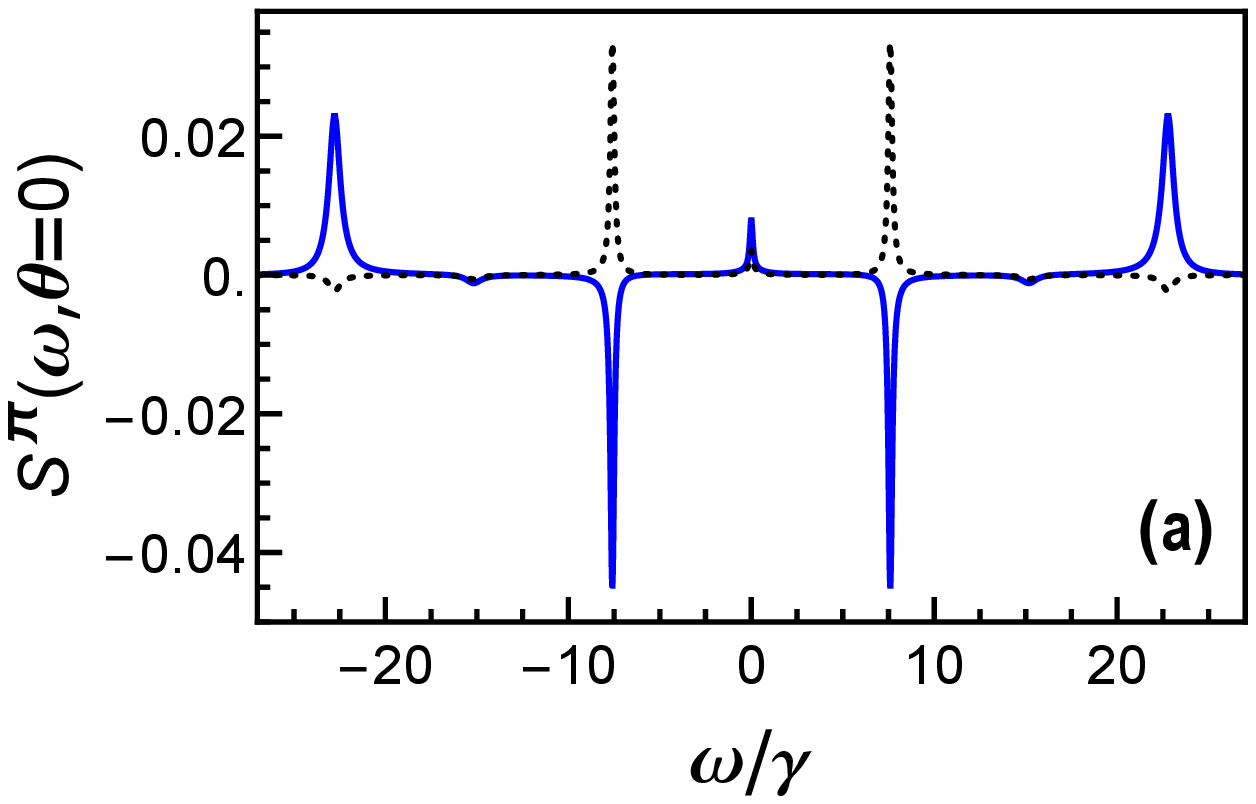}
	\end{center}
	\begin{center}
		\includegraphics[width=8cm,height=5.5cm]{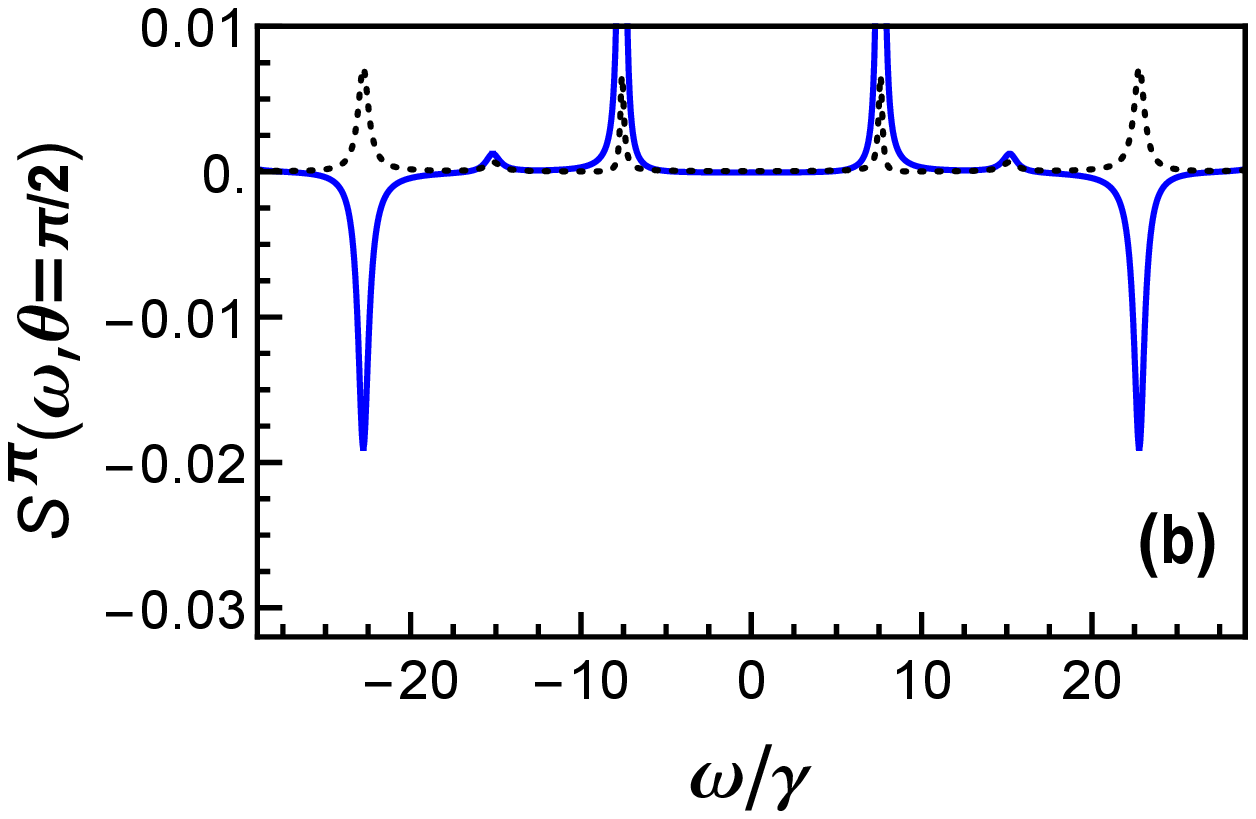}
	\end{center}
	\vspace{-2em}
	\caption{Squeezing spectra $S^{\pi}(\omega, \theta)$ as a function of $\omega$ for strong off-resonance excitation.
The parameters are $\gamma=1$, $\Delta=15$, $\Omega_{a}=4$, $\Omega_{b}=12$, and (a) $\theta=0$ and (b) $\theta=\pi/2$. The solid
(dotted) curves are in the presence (absence) of VIC terms. \vspace{-0.5em}}
\end{figure}
An interpretation of these numerical results can be given using the dressed-state description of the atom-field interaction. The dressed states
$\ket{\mf{\Phi}}$ ($\mf{\Phi}=\alpha,\beta,\kappa,\mu$), which are defined as the eigenstates
($H_I\ket{\mf{\Phi}}=\hbar\lambda_{\mf{\Phi}}\ket{\mf{\Phi}}$) of the interaction Hamiltonian (\ref{hamil}), can be obtained in the basis of the
bare atomic states as
\vspace{-1em}
\begin{align}
\ket{\alpha}=&~c_{\alpha1}\ket{1}+c_{\alpha2}\ket{2}+c_{\alpha3}\ket{3}+c_{\alpha4}\ket{4},\nonumber\\
\ket{\beta}=&~c_{\beta1}\ket{1}+c_{\beta2}\ket{2}+c_{\beta3}\ket{3}+c_{\beta4}\ket{4},\nonumber \\
\ket{\kappa}=&~c_{\kappa1}\ket{1}+c_{\kappa2}\ket{2}+c_{\kappa3}\ket{3}+c_{\kappa4}\ket{4},\nonumber\\
\ket{\mu}=&~c_{\mu1}\ket{1}+c_{\mu2}\ket{2}+c_{\mu3}\ket{3}+c_{\mu4}\ket{4}, \label{dress}
\end{align}
where the expansion coefficients are given by
\begin{eqnarray}
c_{i1}&=&\mc{N}\lambda_{i},~~ c_{i2}=\mc{N}\left[\frac{\lambda_{i}\Omega_{a}\Omega_{b}}{\lambda_{i}
(\Delta+\lambda_{i})-\Omega_{a}^2}\right],\nonumber\\
c_{i3}&=&\mc{N}\Omega_{a},~c_{i4} = \mc{N} \left[\frac{-\lambda_{i}\Omega_{b}(\Delta+\lambda_{i})}
{\lambda_{i}(\Delta+\lambda_{i})-\Omega_{a}^2}\right].\label{cval}
\end{eqnarray}
Here, the overall constant factor $\mc{N}$ is fixed by the normalization condition
$c_{i1}^2+c_{i2}^2+c_{i3}^2+c_{i4}^2=1$. The eigenvalues of the dressed states (\ref{dress}) are
\begin{flalign}
&\lambda_{\alpha}=-\frac{(\sqrt{\Delta^{2}+\Omega_{1}^{2}}+\Delta)}{2}, &\lambda_{\beta}&=-\frac{(\sqrt{\Delta^{2}+\Omega_{2}^{2}}
+\Delta)}{2}, \nonumber\\
&\lambda_{\kappa}=\frac{(\sqrt{\Delta^{2}+\Omega_{2}^{2}}-\Delta)}{2}, &\lambda_{\mu}&=\frac{(\sqrt{\Delta^{2}+\Omega_{1}^{2}}-\Delta)}{2},
\end{flalign}
where the effective Rabi frequencies are $\Omega_{1}=\sqrt{4\Omega_{a}^2+\Omega_{b}^2}+\Omega_{b}$ and
$\Omega_{2}=\sqrt{4\Omega_{a}^2+\Omega_{b}^2}-\Omega_{b}$. The spectral features in the squeezing spectrum can be understood in terms of transitions
between the dressed states $\ket{i}\rightarrow\ket{j} (i,j=\alpha,\beta,\kappa,\mu)$. The transitions between adjacent manifolds of the same
dressed-states give rise to the central peak at $\omega=0$. The sidebands in the spectrum occur at frequencies
$\pm\omega_{ij}=\pm(\lambda_{i}-\lambda_{j})$ as a result of transitions between different dressed states
$\ket{i}\leftrightarrow\ket{j} (i\ne j)$.

In the strong-field limit ($\Omega_{a},\Omega_{b}>>\gamma$), the squeezing spectrum (\ref{pisq}) for the $\pi$-fluorescence field
can be worked out in the dressed-state basis (\ref{dress}). In order to understand the role of VIC in the spectral features, we derive an
analytical formula for the spectrum of the sidebands exhibiting squeezing in figure 4. For the parameters of figure 4, the numerical values
of the eigenvalues (in units of $\gamma$) are $\lambda_{\alpha}=-22.69$, $\lambda_{\beta}=-15.10$, $\lambda_{\kappa}=0.10$,
$\lambda_{\mu}=7.69$. The squeezing (negative peaks) in the inner sidebands of the spectrum $S^{\pi}(\omega,\theta=0)$
(solid curve in figure 4(a)) can be seen as originating from the dressed-state transitions $\ket{\mu}\leftrightarrow\ket{\kappa}$ and
$\ket{\beta}\leftrightarrow\ket{\alpha}$. Similarly, the transitions $\ket{\mu}\leftrightarrow\ket{\beta}$ and
$\ket{\kappa}\leftrightarrow\ket{\alpha}$ of the dressed states contribute to the squeezing in the outer sidebands of the spectrum
$S^{\pi}(\omega,\theta=\pi/2)$ (solid curve in figure 4(b)). Considering these atomic transitions, the spectra at the sidebands can be
obtained in the dressed-state representation as
\begin{flalign}
&S^{\pi}(\omega_{\pm},0)=\frac{\mc{W}^{\pi}_{+}\,\,\Gamma_{+}}{[\Gamma_{+}^2+(\omega\mp\omega_{\mu\kappa})^2]}+
\frac{\mc{W}^{\pi}_{-}\,\,\Gamma_{-}}{[\Gamma_{-}^2+(\omega\mp\omega_{\mu\kappa})^2]},& \nonumber\\
&S^{\pi}(\omega_{\pm},\pi/2)=\frac{\tilde{\mc{W}}^{\pi}_{+}\,\,\tilde{\Gamma}_{+}}{[\tilde{\Gamma}_{+}^2+(\omega\mp
\omega_{\mu\beta})^2]}+\frac{\tilde{\mc{W}}^{\pi}_{-}\,\,\tilde{\Gamma}_{-}}{[\tilde{\Gamma}_{-}^2+(\omega\mp\omega_{\mu\beta})^2]}.
\label{drespi} &
\end{flalign}
In the above equations, the upper [lower] sign is for the positive $(\omega>0)$ [negative $(\omega<0)$] part of the spectrum along the
$\omega$-axis. To derive these sideband spectra, we have included the VIC-term $(\gamma_{12}=-\gamma/3)$ in the calculations. Equations
(\ref{drespi}) show that two different Lorentzians contribute to each of the sidebands centered at $\omega=\pm\omega_{\mu\kappa}$
and $\omega=\pm\omega_{\mu\beta}$ in the spectra $S^{\pi}(\omega,0)$ and $S^{\pi}(\omega,\pi/2)$ (see figure 4), respectively. The
explicit forms of the widths $(\Gamma_{\pm},\tilde{\Gamma}_{\pm})$ and weights $(\mc{W}^{\pi}_{\pm},\tilde{\mc{W}}^{\pi}_{\pm})$ of
the Lorentzians in equations (\ref{drespi}) are given in Appendix. On substituting the expansion coefficients (\ref{cval}) into
the expressions for $\Gamma_{\pm},\tilde{\Gamma}_{\pm},\mc{W}^{\pi}_{\pm},\tilde{\mc{W}}^{\pi}_{\pm}$ (see Appendix), the
formulas (\ref{drespi}) provide good agreement with the squeezing peaks (negative peaks) shown in figure 4.
\begin{figure}[b]
	\begin{center}
		\includegraphics[width=8cm,height=5.5cm]{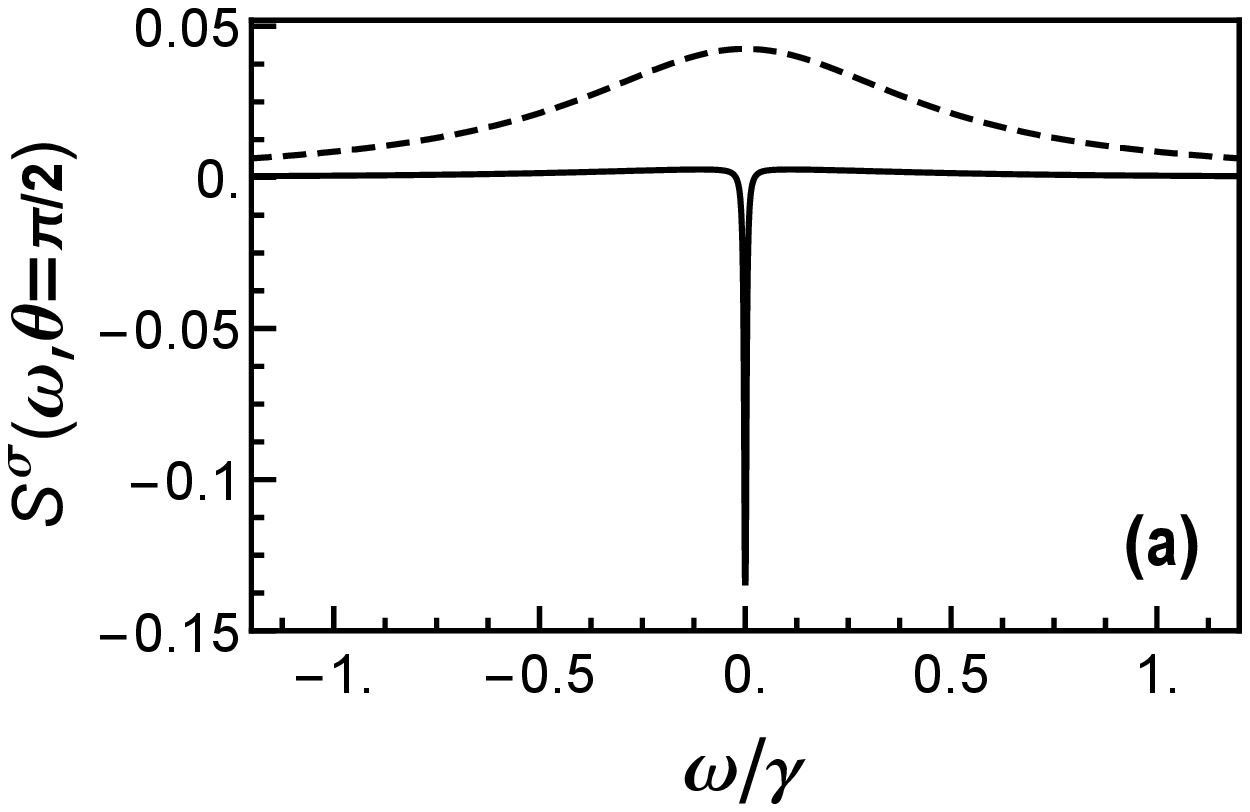}
	\end{center}
	\begin{center}
		\includegraphics[width=8cm,height=5.5cm]{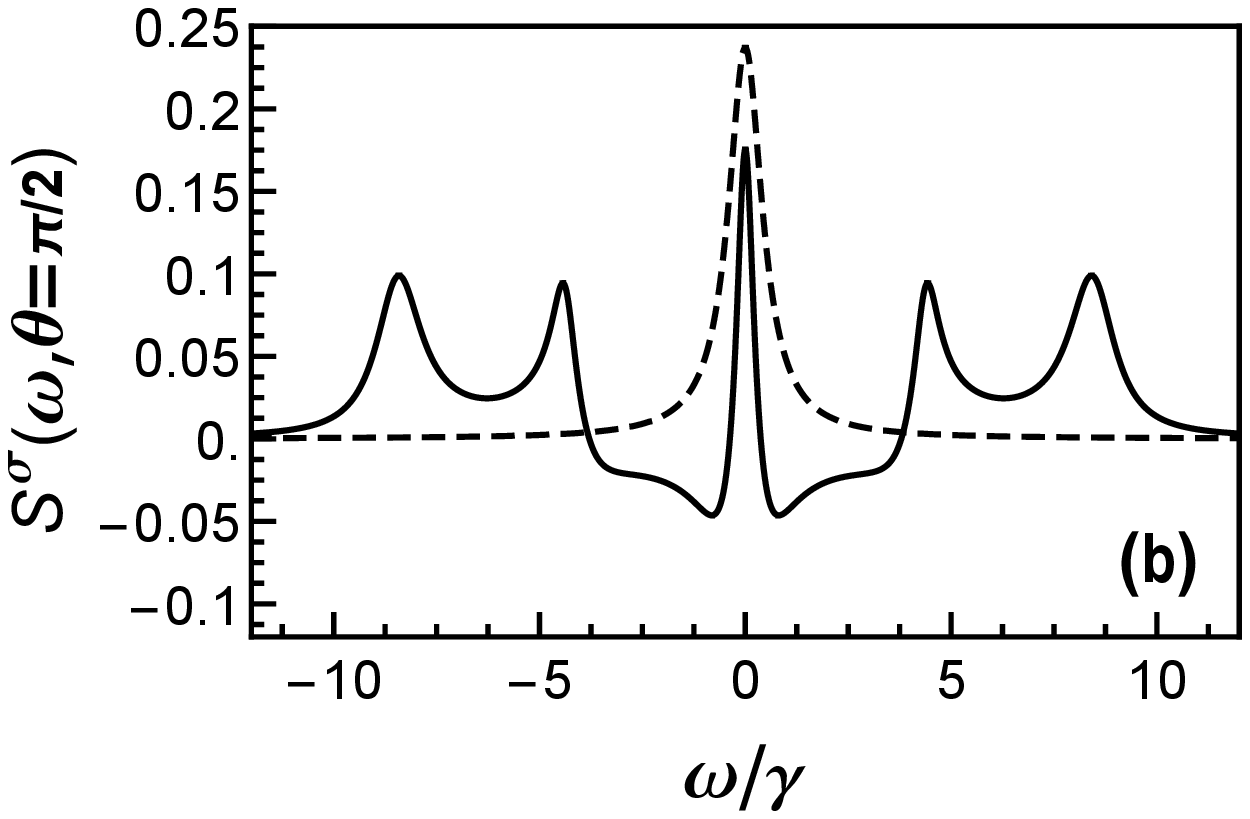}
	\end{center}
	\vspace{-1.5em}
	\caption{Influence of the additional field on the squeezing spectra  $S^{\sigma}(\omega, \theta)$ as a function of $\omega$.
The parameters are $\theta=\pi/2$, $\gamma=1$, $\phi=0$, $\Delta=0$, and (a) $\Omega_{a}=\Omega_{b}=0.02$ and (b) $\Omega_{a}=1$,
$\Omega_{b}=4$. The dashed curves represent the spectra for $\Omega_{b}=0$ with other parameters remaining same. Actual values of the
dashed curve are 0.1 [5] times that shown in graph (a) [(b)].\vspace{-1em}}
\end{figure}
\vspace{-1em}
\subsection{Squeezing spectrum - $\sigma$ transitions}
\vspace{-0.5em}
\subsubsection{Effect of the additional field}
\vspace{-1em}
We now proceed to analyze the squeezing spectrum $S^{\sigma}(\omega,\theta)$ of the fluorescence field emitted on the $\sigma$ transitions.
While the squeezing spectrum of the $\pi$-fluorescence has been studied by Tan \textit{et al.} \cite{tan} in the absence of the additional field
$(\Omega_b=0)$, that for $\sigma$-fluorescence has not been reported so far. We show here that the $\sigma$-fluorescence displays interesting
squeezing characteristics that distinguish it from $\pi$-fluorescence. In the absence of the circularly polarized light $(\Omega_b=0)$, one
can easily verify that when considering $\theta=\pi/2$ (out-of-phase quadrature) and zero detuning $(\Delta=0)$, the squeezing spectrum of
the $\sigma$ transitions (\ref{sigsq}) is identical to that of the in-phase quadrature $(\theta=0)$ of a driven two-level atom with Rabi
frequency $\Omega=2\Omega_{a}$. It is well known that the in-phase quadrature of the two-level atom does not exhibit squeezing for
resonance excitations \cite{zoller}. However, the situation is different when the additional field drives the atom.
	
In figure 5 we show the squeezing spectra (\ref{sigsq}) of the out-of-phase quadrature when considering both weak and moderately strong
driving fields on resonance. For the cases when $\Omega_{b}=0$, the squeezing is absent (see dashed curves in figure 5) in the
spectra similar to that in the in-phase quadrature spectra of a driven two-level atom. In contrast, an ultranarrow squeezing peak appears
at the laser frequency in the presence of a weak additional field $(\Omega_{b}\ll\gamma)$ as shown by the solid curve in figure 5(a). For moderate
driving field strengths, there is no squeezing at the laser frequency, whereas the spectral squeezing gets shifted to the wing portion of the
central peak (see solid curve in figure 5(b)).

\begin{figure}[t]
	\begin{center}
		\includegraphics[width=8cm,height=5.5cm]{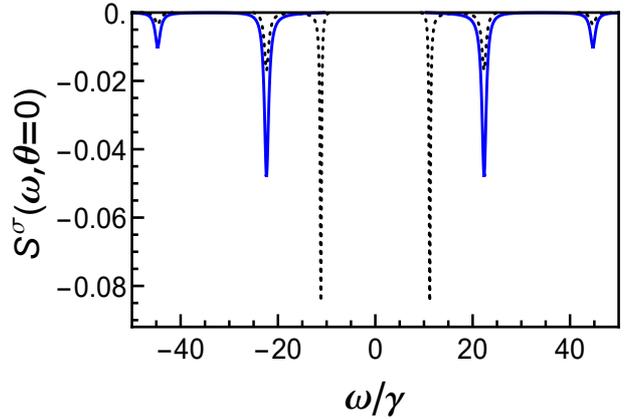}
	\end{center}
	\vspace{-2em}
	\caption{Squeezing spectra $S^{\sigma}(\omega, \theta)$ as a function of $\omega$ for strong off-resonance excitation. The parameters
are $\theta=0$, $\gamma=1$, $\Delta=20$, $\Omega_{a}=10$, $\Omega_{b}=15$. The solid (dotted) curve is for $\phi = \pi/2$ ($\phi=0$). The scale
on the graphs is chosen to show only the squeezing peaks in the spectrum.}
\vspace{-1em}
\end{figure}
\vspace{-2em}
\subsubsection{Effect of the relative phase}
\vspace{-1em}
When the additional field is not applied $(\Omega_b=0)$ on the atomic system, the squeezing properties of the fluorescence field do not depend on the
phase of the driving field $(\Omega_a)$ acting on the $\pi$ transitions \cite{tan}. However, the presence of the additional field $(\Omega_b\neq0)$
driving the system brings a relative phase $(\phi)$ dependence in the squeezing spectrum (\ref{sigsq}) of the $\sigma$ transitions.
We proceed to show how the relative phase modifies the squeezing properties of the $\sigma$-fluorescence field. Phase control of squeezing has
been reported earlier in the literature \cite{anton1, arun11}. In these works, the phase control was achieved either through vacuum-induced coherences
or by using a closed-loop scheme of atomic transitions. It was shown that by changing the relative phases of the driving fields the squeezing
could be suppressed or canceled \cite{anton1} and also be shifted from inner- to outer-sidebands of the spectrum \cite{arun11}.
In the present system, the phase control of spectral features appears because of the polarization-dependent detection schemed used to observe
the fluorescence light as shown in figure 1(b). Note that the phase dependence occurs because only the $\mf{e}_{x}$-component of the fluorescence
field is detected along the $\sigma$ transitions \cite{heb}. To demonstrate the role of the relative phase, we plot in figure 6 the in-phase
quadrature spectrum for strong off-resonance excitation. For relative phase $\phi=0$ (dotted curve in figure 6), it is seen that the squeezing
appears in six peaks. As the relative phase is changed to $\phi=\pi/2$, the squeezing in the inner sidebands near the central region is cancelled,
whereas the squeezing in other peaks gets enhanced (solid line in figure 6). Thus, one finds that the relative phase $(\phi)$ can not only
suppress squeezing as shown in previous studies \cite{anton1} but also cause enhancement of squeezing in the spectrum at the same time. This
result is closely similar to that in the incoherent spectrum of resonance fluorescence reported in our previous work \cite{heb}.

The enhancement and suppression of the squeezing peaks can be explained using the dressed-state picture. In order to understand the
phase-dependent spectral features shown in figure 6, we work out the analytical formula for the spectrum (\ref{sigsq}) in the
dressed-state formalism as \cite{phinote}
\begin{flalign}
S^{\sigma}(\omega_{\pm},0)=&\frac{\mc{W}^{\sigma}_{1}\,\,\Gamma_{1}}{[\Gamma_{1}^2+(\omega\mp\omega_{\mu\alpha})^2]}+\frac{\mc{W}^{\sigma}_{2}\,\,
\Gamma_{2}}{[\Gamma_{2}^2+(\omega\mp\omega_{\kappa\beta})^2]}&\nonumber\\&+\frac{\mc{W}^{\sigma}_{+}\,\,\Gamma_{+}}{[\Gamma_{+}^2+
(\omega\mp\omega_{\mu\kappa})^2]}+\frac{\mc{W}^{\sigma}_{-}\,\,\Gamma_{-}}{[\Gamma_{-}^2+(\omega\mp\omega_{\mu\kappa})^2]},&\label{dresig}
\end{flalign}
where the upper (lower) sign is for the squeezing peaks on the positive (negative) portion of the $\omega$-axis.
In the above equation, the first two terms are contributions due to the single dressed-state transitions
$\ket{\mu}\leftrightarrow\ket{\alpha}$ and $\ket{\kappa}\leftrightarrow\ket{\beta}$, respectively, whereas the last two terms originate from
the coupled transitions $\ket{\mu}\leftrightarrow\ket{\kappa}$ and $\ket{\beta}\leftrightarrow\ket{\alpha}$. The expressions for the widths
$(\Gamma_{1},\Gamma_{2},\Gamma_{\pm})$ and weights $(\mc{W}^{\sigma}_{1},\mc{W}^{\sigma}_{2},\mc{W}^{\sigma}_{\pm})$ in
equation (\ref{dresig}) are given in Appendix. From the analytical expressions (see Appendix) for the weights
$(\mc{W}^{\sigma}_{1},\mc{W}^{\sigma}_{2}, \mc{W}^{\sigma}_{\pm})$ in equation (\ref{dresig}), it is found that the squeezing peaks
at $\omega=\pm\omega_{\mu\alpha}$ and $\omega=\pm\omega_{\kappa\beta}$ get enhanced as the phase $\phi$ is changed
from $0$ to $\pi/2$, whereas the sidebands at $\omega=\pm\omega_{\mu\kappa}$ are suppressed for $\phi=\pi/2$ as shown in figure 6.

\vspace{-1em}
\section{\label{sec:concl}conclusions}
\vspace{-1em}
In conclusion, we have investigated theoretically the squeezing properties of the fluorescence radiation from a $J=1/2$ to $J=1/2$
system driven by two coherent fields. Specifically, we have studied the effects of VIC in the squeezing spectrum of the fluorescence
emitted on $\pi$ transitions in the atom. It is found that VIC induces spectral squeezing in the fluorescence of $\pi$ transitions
for weak as well as strong driving fields. The origin of spectral squeezing in the $\pi$-fluorescence has been explained using a
dressed-state analysis of the atom-field interaction. The squeezing spectrum of the fluorescence field emitted along the
$\sigma$ transitions is also investigated. It has been shown that the squeezing spectrum of the $\sigma$-fluorescence field exhibits
a strong dependence on the relative phase of the driving fields even though the atomic population dynamics is phase-independent.
In particular, the squeezing peaks can be either enhanced or suppressed by adjusting the relative phases of the driving fields.

\vspace{-1em}
\appendix*
\section{calculation of the widths and weights of the spectral lines}
\vspace{-0.5em}
\subsection{The decay rates in the dressed-state picture}
\vspace{-1em}
In the secular approximation, the equations of motion for the dressed-state coherences are given by
\begin{equation}
\dot{\rho}_{\mu\alpha}= -(\Gamma_{1}+i\omega_{\mu\alpha})\rho_{\mu\alpha}, \nonumber
\end{equation}
\begin{equation}
\dot{\rho}_{\kappa\beta}= -(\Gamma_{2}+i\omega_{\kappa\beta})\rho_{\kappa\beta}, \nonumber
\end{equation}
\begin{equation}
\dot{\rho}_{\mu\beta}= -(\Gamma_{3}+i\omega_{\mu\beta})\rho_{\mu\beta}+\Gamma_{4}\rho_{\kappa\alpha},
\nonumber
\end{equation}
\begin{equation}
\dot{\rho}_{\kappa\alpha}= -(\Gamma_{5}+i\omega_{\kappa\alpha})\rho_{\kappa\alpha}+\Gamma_{6}\rho_{\mu\beta},
\nonumber
\end{equation}
\begin{equation}
\dot{\rho}_{\mu\kappa}= -(\Gamma_{7}+i\omega_{\mu\kappa})\rho_{\mu\kappa}+\Gamma_{8}\rho_{\beta\alpha},
\nonumber
\end{equation}
\begin{equation}
\dot{\rho}_{\beta\alpha}= -(\Gamma_{9}+i\omega_{\beta\alpha})\rho_{\beta\alpha}+\Gamma_{10}\rho_{\mu\kappa},
\label{dreseq}
\end{equation}
where
\begin{flalign}
\Gamma_{1}=&~\frac{\gamma}{6}[4(c_{\alpha1}^2c_{\alpha3}^2+c_{\alpha2}^2c_{\alpha3}^2+c_{\alpha1}^2
c_{\alpha4}^2)+3]&\nonumber\\&-2\gamma_{12}c_{\alpha1}^2c_{\alpha3}^2,&
\label{gterm}
\end{flalign}
\begin{flalign}
\Gamma_{2}=&~\frac{\gamma}{6}[4(c_{\beta1}^2c_{\beta3}^2+c_{\beta2}^2c_{\beta3}^2+c_{\beta1}^2
c_{\beta4}^2)+3]&\nonumber\\&-2\gamma_{12}c_{\beta1}^2c_{\beta3}^2,&
\label{gterm2}
\end{flalign}
\begin{flalign}
\Gamma_{3}=&~\frac{\gamma}{6}[3(c_{\beta1}^2+c_{\beta2}^2+c_{\alpha3}^2+c_{\alpha4}^2)-4c_{\alpha1}
c_{\alpha3}c_{\beta1}c_{\beta3}]&\nonumber\\&-2\gamma_{12}c_{\alpha1}c_{\alpha3}c_{\beta1}c_{\beta3},&
\label{gterm3}
\end{flalign}
\begin{flalign}
\Gamma_{4}=&~\frac{2\gamma}{3}[c_{\alpha1}^2c_{\beta4}^2+c_{\alpha2}^2c_{\beta3}^2-c_{\alpha1}c_{\alpha2}
c_{\beta3}c_{\beta4}]&\nonumber\\&-\gamma_{12}(c_{\alpha1}^2c_{\beta3}^2+c_{\alpha2}^2c_{\beta4}^2),&
\label{gterm4}
\end{flalign}
\begin{flalign}
\Gamma_{5}=&~\frac{\gamma}{6}[3(c_{\alpha1}^2+c_{\alpha2}^2+c_{\beta3}^2+c_{\beta4}^2)-4c_{\alpha1}
c_{\alpha3}c_{\beta1}c_{\beta3}]&\nonumber\\&-2\gamma_{12}c_{\alpha1}c_{\alpha3}c_{\beta1}c_{\beta3},&
\label{gterm5}
\end{flalign}
\begin{flalign}
\Gamma_{6}=&~\frac{2\gamma}{3}[c_{\alpha3}^2c_{\beta2}^2+c_{\alpha4}^2c_{\beta1}^2-c_{\alpha3}
c_{\alpha4}c_{\beta1}c_{\beta2}]&\nonumber\\&-\gamma_{12}(c_{\alpha3}^2c_{\beta1}^2+c_{\alpha4}^2
c_{\beta2}^2),&
\label{gterm6}
\end{flalign}
\begin{flalign}
\Gamma_{7}=&~\frac{\gamma}{6}[3(c_{\alpha3}^2+c_{\alpha4}^2+c_{\beta3}^2+c_{\beta4}^2)+4c_{\alpha1}
c_{\alpha3}c_{\beta1}c_{\beta3}]&\nonumber\\&+2\gamma_{12}c_{\alpha1}c_{\alpha3}c_{\beta1}c_{\beta3},&
\label{gterm7}
\end{flalign}
\begin{flalign}
\Gamma_{8}=&-\frac{\gamma}{6}[4c_{\alpha1}^2c_{\beta1}^2+3(c_{\alpha4}c_{\beta1}-c_{\alpha3}c_{\beta2})^2&
\nonumber\\
&+3(c_{\alpha2}c_{\beta3}-c_{\alpha1}c_{\beta4})^2]+\gamma_{12}(c_{\alpha1}^2c_{\beta2}^2+c_{\alpha2}^2
c_{\beta1}^2),&
\label{gterm8}
\end{flalign}
\begin{flalign}
\Gamma_{9}=&~\frac{\gamma}{6}[3(c_{\alpha1}^2+c_{\alpha2}^2+c_{\beta1}^2+c_{\beta2}^2)+4c_{\alpha1}
c_{\alpha3}c_{\beta1}c_{\beta3}]&\nonumber\\
&+2\gamma_{12}c_{\alpha1}c_{\alpha3}c_{\beta1}c_{\beta3},&
\label{gterm9}
\end{flalign}
\begin{flalign}
\Gamma_{10}=&~-\frac{\gamma}{6}[4c_{\alpha3}^2c_{\beta3}^2+3(c_{\alpha4}c_{\beta1}-c_{\alpha3}c_{\beta2})^2&
\nonumber\\
&+3(c_{\alpha2}c_{\beta3}-c_{\alpha1}c_{\beta4})^2]+\gamma_{12}(c_{\alpha3}^2c_{\beta4}^2+c_{\alpha4}^2
c_{\beta3}^2).&
\label{gterm10}
\end{flalign}
The effective decay rates in equations (\ref{drespi}) and (\ref{dresig}) are
\begin{flalign}
\Gamma_{\pm}=&~[-(\Gamma_{7}+\Gamma_{9})\pm\sqrt{(\Gamma_{7}-\Gamma_{9})^2+4\Gamma_{8}\Gamma_{10}}]/2,&\nonumber\\
\tilde{\Gamma}_{\pm}=&~[-(\Gamma_{3}+\Gamma_{5})\pm\sqrt{(\Gamma_{3}-\Gamma_{5})^2+4\Gamma_{4}\Gamma_{6}}]/2.&\label{effdec}
\end{flalign}
In deriving the decay rates (\ref{gterm})-(\ref{gterm10}) of the dressed-state coherences, we use the following relations
among the expansion coefficients (\ref{cval}):
\begin{flalign}
&c_{\mu1}=-c_{\alpha4},~~c_{\mu2}=c_{\alpha3},~~c_{\mu3}=-c_{\alpha2},~~c_{\mu4}=c_{\alpha1},&\nonumber\\
&c_{\kappa1}=c_{\beta4},~~c_{\kappa2}=-c_{\beta3},~~c_{\kappa3}=c_{\beta2},~~c_{\kappa4}=-c_{\beta1},&
\nonumber \\
&c_{\alpha2}c_{\alpha4}=-c_{\alpha1}c_{\alpha3},~~c_{\beta2}c_{\beta4}=-c_{\beta1}c_{\beta3},&\nonumber\\
&c_{\alpha2}c_{\beta2}=-c_{\alpha1}c_{\beta1},~~c_{\alpha4}c_{\beta4}=-c_{\alpha3}c_{\beta3}.\label{crel}&
\end{flalign}

\vspace{-1em}
\subsection{The weights of the spectral lines in $\pi$-fluorescence squeezing spectrum}
\vspace{-1em}
The weights of the sidebands in the spectrum (\ref{drespi}), using equations (\ref{crel}),
can be derived as
\begin{flalign}
\mc{W}^{\pi}_{\pm}=&~\frac{\gamma}{6(\Gamma_{-}-\Gamma_{+})}(c_{\alpha1}c_{\beta3}+c_{\alpha3}
c_{\beta1}-c_{\alpha2}c_{\beta4}-c_{\alpha4}c_{\beta2})\nonumber\\
&~\times[(c_{\alpha1}c_{\beta3}-c_{\alpha2}c_{\beta4})F_{\pm}+(c_{\alpha3}c_{\beta1}-c_{\alpha4}
c_{\beta2})G_{\pm}],
\end{flalign}
\begin{flalign}
\tilde{\mc{W}}^{\pi}_{\pm}=&~\frac{\gamma}{6(\tilde{\Gamma}_{-}-\tilde{\Gamma}_{+})}(c_{\alpha2}
c_{\beta1}+c_{\alpha1}c_{\beta2}-c_{\alpha4}c_{\beta3}-c_{\alpha3}c_{\beta4})\nonumber\\
&~\times[(c_{\alpha2}c_{\beta1}+c_{\alpha1}c_{\beta2})\tilde{F}_{\pm}-(c_{\alpha4}c_{\beta3}+
c_{\alpha3}c_{\beta4})\tilde{G}_{\pm}],
\end{flalign}
where
\begin{flalign}
F_{\pm}=&\pm(\Gamma_{7}-\Gamma_{9}+2\Gamma_{8})\rho_{\alpha\alpha}\pm(\Gamma_{9}-\Gamma_{7}+
2\Gamma_{10})\rho_{\kappa\kappa}&\nonumber\\
&+(\Gamma_{+}-\Gamma_{-})(\rho_{\alpha\alpha}+\rho_{\kappa\kappa}),&\nonumber\\
G_{\pm}=&\pm(\Gamma_{7}-\Gamma_{9}+2\Gamma_{8})\rho_{\beta\beta}\pm(\Gamma_{9}-\Gamma_{7}+
2\Gamma_{10})\rho_{\mu\mu}&\nonumber\\
&+(\Gamma_{+}-\Gamma_{-})(\rho_{\beta\beta}+\rho_{\mu\mu}),& \nonumber \\
\tilde{F}_{\pm}=&\pm(\Gamma_{3}-\Gamma_{5}-2\Gamma_{4})\rho_{\alpha\alpha}\pm(\Gamma_{5}-
\Gamma_{3}-2\Gamma_{6})\rho_{\beta\beta}&\nonumber\\
&+(\tilde{\Gamma}_{+}-\tilde{\Gamma}_{-})(\rho_{\alpha\alpha}+\rho_{\beta\beta}),&\nonumber\\
\tilde{G}_{\pm}=&\pm(\Gamma_{3}-\Gamma_{5}-2\Gamma_{4})\rho_{\kappa\kappa}\pm(\Gamma_{5}-
\Gamma_{3}-2\Gamma_{6})\rho_{\mu\mu}&\nonumber\\
&+(\tilde{\Gamma}_{+}-\tilde{\Gamma}_{-})(\rho_{\kappa\kappa}+\rho_{\mu\mu}).& \label{cons}
\end{flalign}
\vspace{-1em}
\subsection{The weights of the spectral lines in $\sigma$-fluorescence squeezing spectrum}
\vspace{-1em}
The weights of the sidebands in the spectrum (\ref{dresig}), using equations (\ref{crel}),
can be derived as
\begin{flalign}
\mc{W}^{\sigma}_{1}=&~\gamma_{\sigma}\big\{(c_{\alpha1}^4+c_{\alpha2}^4)\rho_{\alpha\alpha}+
(c_{\alpha3}^4+c_{\alpha4}^4)\rho_{\mu\mu}&\nonumber\\
&-c_{\alpha1}^2c_{\alpha4}^2(\rho_{\alpha\alpha}+\rho_{\mu\mu})-2\cos2\phi[c_{\alpha1}^2
c_{\alpha2}^2\rho_{\alpha\alpha}&\nonumber\\&+c_{\alpha3}^2c_{\alpha4}^2\rho_{\mu\mu}-
c_{\alpha1}^2c_{\alpha3}^2(\rho_{\alpha\alpha}+\rho_{\mu\mu})]&\nonumber\\
&-\cos4\phi [c_{\alpha2}^2c_{\alpha3}^2(\rho_{\alpha\alpha}+\rho_{\mu\mu})]\big\},&\label{wtsig1}
\end{flalign}
\begin{flalign}
\mc{W}^{\sigma}_{2}=&~\gamma_{\sigma}\big\{(c_{\beta1}^4+c_{\beta2}^4)\rho_{\beta\beta}+
(c_{\beta3}^4+c_{\beta4}^4)\rho_{\kappa\kappa}&\nonumber\\
&-c_{\beta1}^2c_{\beta4}^2(\rho_{\beta\beta}+\rho_{\kappa\kappa})-2\cos2\phi[c_{\beta1}^2
c_{\beta2}^2\rho_{\beta\beta}&\nonumber\\
&+c_{\beta3}^2c_{\beta4}^2\rho_{\kappa\kappa}-c_{\beta1}^2c_{\beta3}^2(\rho_{\beta\beta}+
\rho_{\kappa\kappa})]&\nonumber\\
&-\cos4\phi [c_{\beta2}^2c_{\beta3}^2(\rho_{\beta\beta}+\rho_{\kappa\kappa})]\big\},&\label{wtsig2}
\end{flalign}
\begin{flalign}
\mc{W}^{\sigma}_{\pm}=&~\frac{\gamma_{\sigma}}{2(\Gamma_{-}-\Gamma_{+})}\times\nonumber\\
&~\big\{(c_{\alpha1}^2c_{\beta4}^2+c_{\alpha2}^2c_{\beta3}^2+2\cos2\phi \,c_{\alpha1}c_{\alpha2}
c_{\beta3}c_{\beta4})\,F_{\pm}&\nonumber\\
&+(c_{\alpha3}^2c_{\beta2}^2+c_{\alpha4}^2 c_{\beta1}^2+2\cos2\phi \,c_{\alpha3}c_{\alpha4}
c_{\beta1}c_{\beta2})\,G_{\pm}&\nonumber\\
&-4\cos^2\phi\cos2\phi \,\, c_{\alpha1}c_{\alpha3}c_{\beta1}c_{\beta3}(F_{\pm}+G_{\pm})\big\},&\label{wtsigpm}
\end{flalign}
where $F_{\pm}$ and $G_{\pm}$ are given in equations (\ref{cons}).


\begin{thebibliography}{56}
\bibitem{phinote} The exponential phase factors in equation (\ref{sigsq}) are assumed to be real in
the analytical calculations, i.e., $e^{\pm2i\phi}=\cos(2\phi)$ and $e^{-4i\phi}=\cos(4\phi)$.
\end{thebibliography}
\end{document}